\documentclass[pdflatex,sn-standardnature]{sn-jnl}

\usepackage{amsmath}
\usepackage{lineno}
\usepackage{natbib}
\usepackage{mathtools}
\usepackage[section]{placeins}
\usepackage{tabularx}

\jyear{2024}%

\theoremstyle{thmstyleone}%
\theoremstyle{thmstyletwo}%

\theoremstyle{thmstylethree}%

\begin{document}

\title[A New Thermodynamic Approach to Multimode Fibre SC and Soliton Cond.]{A New Thermodynamic Approach to Multimode Fibre Self-cleaning and Soliton Condensation}


\author*[1]{\fnm{Mario} \sur{Zitelli}}\email{mario.zitelli@uniroma1.it}

\affil*[1]{\orgdiv{Department of Information Engineering, Electronics and Telecommunications}, \orgname{Università degli Studi di Roma Sapienza}, \orgaddress{\street{Via Eudossiana 18}, \city{Rome}, \postcode{00184}, \state{RM}, \country{Italy}}}

\abstract{A new thermodynamic theory for optical multimode systems is proposed. Theory is based on a weighted Bose-Einstein law, and includes the state equation, the fundamental equation for the entropy and a metric to measure the accuracy of the thermodynamic approach. The theory is used to compare the experimental results of two propagation regimes in multimode fibres, specifically the self-cleaning in the normal chromatic dispersion region and the soliton condensation in the anomalous dispersion region. Surprising similarities are found in terms of thermodynamic parameters, suggesting a common basis for the thermalisation processes observed in the two propagation regimes.}


\maketitle

\section{Introduction} \label{sec:Intro}

Multimode optical structures have been the subject of intense studies in the last decade. Multicore and multimode fibres, proposed half a century ago \cite{Gloge:6767859,Hasegawa:80}, have been re-discovered to solve the problem of capacity crunch in optical transport networks \cite{Richardson-NatPhot-2013-94-2013,Winzer:18,Essiambre:5420239}, or to generate high quality power beams \cite{Wright2017}. Photonic lattice have been proposed as flatbands for applications as slow light or novel photonic crystal fibres \cite{10.1063/1.5034365}. Multimode fibres were also proposed for high-resolution nonlinear fluorescence imaging \cite{moussa2021spatiotemporal}.

Under suitable conditions, multimodal systems can give rise to thermalisation processes, in which power distribution between system modes evolves irreversibly towards the thermodynamic equilibrium state of maximum Boltzmann entropy. In the weakly nonlinear regime, this process was initially described by the wave turbulence theory \cite{Zakharov1992,PICOZZI20141}.

In one of the first attempts to describe a complex optical system in thermodynamic terms \cite{Markvart_2008}, the étendue of a light beam upon absorption and re-emission was compared to the thermodynamics of a two-dimensional gas. In \cite{PhysRevLett.101.143901}, the emission spectra of colloidal dye-doped laser was experimentally and theoretically analyzed, finding analogies with the Gross-Pitaevskii models for Bose-Einstein condensates.
Experiments of beam self-cleaning in the normal dispersion region, an optical effect related to the inter-modal four-wave mixing (IM-FWM) \cite{Wright2015c}\cite{Renninger2013} and taking place in multimode fibres, where the optical power flows towards the lower-order group of modes, were proposed in \cite{Krupa2017}, but only later \cite{baudin2020classical,BAUDIN2023129716} those effects were analyzed in thermodynamic terms and related to thermalised states of a multimode fibre. 

In \cite{Wu2019} a complete thermodynamic theory of optical multimode systems was introduced, using the Rayleigh-Jeans law (RJ) as the basic model for describing the power distribution among the system's modes; theory was developed for microcanonical ergodic systems with constant power $P$ and internal energy $U$. The theory was applied to different multimode systems, such as a multimode graded-index fibre (GRIN), a nonlinear multicore Lieb waveguide lattice and a three-dimensional coupled cavity nonlinear system.
An alternative derivation of the thermodynamic theory, still in terms of RJ law, was found in \cite{Makris:20}; by adopting a grand canonical description of the multimode system, maximisation of the entropy was performed in terms of probability density distribution of the system's microstates, obtaining a Gibbs distribution which was reduced to a RJ for the average mode power occupancy at thermal equilibrium.

The interplay of wave thermalisation and linear disorder, represented by the linear random mode coupling (RMC) of multimode fibres \cite{Gloge:6774107}\cite{Savovi2019PowerFI}, was theoretically and experimentally analyzed in \cite{PhysRevLett.129.063901} and \cite{Fusaro_PhysRevLett.122.123902}; a weak disorder was found to accelerate the rate of thermalisation and condensation. 
In \cite{Zhong2023} thermalised states were numerically calculated under a wide range of nonlinear conditions beyond the IM-FWM. These include second harmonic generation (SHG), multi-wave mixing (MWM) and optomechanical (OM) cascaded interactions between optical and mechanical modes. Steady states similar to thermalisation were experimentally observed in \cite{Zitelli_23}, in pseudo-linear systems composed by long spans of multimode GRIN fibre affected by RMC; output modal distribution were described in terms of a Bose-Einstein law.

Experimental self-cleaning thermalisation was also observed in \cite{pourbeyram2022direct}, propagating 200-fs pulses at $\lambda=1040$ nm wavelength, over few meters (0.5 to 1.5 m) of OM4 GRIN fibre supporting 110 modes per polarization. The GRIN fibre was chosen because its equally spaced propagation constants facilitate IM-FWM. Output field was measured by off-axis digital holography and decomposed into a spectral resolved eigenmode basis. 

A 3D modal decomposition method was applied in \cite{zitelli2023statistics} to analyze the condensation process during soliton propagation in 830 m of GRIN OM4 fibre. 250-fs pulses were propagated at $\lambda=1400$ nm in the anomalous dispersion region of the fibre. Pulses at relatively low power were separated by the modal dispersion, after having interacted by RMC and IM-FWM; for increasing power, a train of dispersive pulses converged to a train of quasi-solitons, and finally to a single soliton condensed to the fundamental mode.

Most of the listed works have relied on the RJ law to describe the distribution under thermalisation conditions. The experimental methods used in \cite{pourbeyram2022direct} and  \cite{zitelli2023statistics}, however, have demonstrated particular accuracy in measuring the power distribution of the higher order modes of a fibre (HOM), allowing a better analysis of the distributions. In this work, we introduce a complete thermodynamic theory based on a weighted Bose-Einstein law (wBE); new expressions for the state equation and the fundamental equation of the Boltzmann entropy will be introduced; an optical Shannon or information entropy will be compared. The theory is tested against the results of the last two experimental works, showing higher accuracy respect to the RJ in reproducing the experimental modal distributions both in the thermalisation regime induced by IM-FWM and in steady state regimes caused by the RMC. Thermodynamic similarities are found between the self-cleaning regime of a GRIN fibre, in the normal chromatic dispersion region, and the multimode soliton condensation process in the anomalous dispersion region, similarities that allow a common thermodynamic description of two operationally different propagation regimes.

\section{Theory} \label{sec:Theory}

Commercial parabolic OM4 GRIN fibres are capable of propagating $Q$ groups of Laguerre-Gauss degenerate modes; each group includes $g_j=2j$ modes and polarizations, $j=1,..,Q$. The total number of propagated modes and polarizations is $2M=Q(Q+1)$; the nearly degenerate modes of a group have substantially equal propagation constant \cite{Olshansky:75} $\beta_j =n_{co} k_0 \sqrt{1-2\Delta (j/Q)}$, with $n_{co}$ the core index and $\Delta=(n_{co}^2-n_{cl}^2)/2n_{co}^2$ the relative core-cladding index difference.

In this work, experiments are considered both in the normal and the anomalous chromatic dispersion regimes with variable input pulse peak power $P$ (W). Correspondingly, it is scaled the power factor $\gamma=N/n_0=P/P_0$, with $N$ the total number of particles in the system, $n_0$ a reference number of particles (see Section Methods), and $P$ and $P_0$ the corresponding input peak powers.

In a generic multimode system, suppose that $n_j$ is the bosonic population into modal group $j$, distributed over $g_j$ nearly-degenerate modes and polarizations. $\epsilon_j=\beta_j-\beta_{j=Q}$ are the differential modal eigenvalues. In experiments where the input power and internal energy scale proportionally, the normalized internal energy $U_N=-\sum_{j}\beta_j n_j/N$ (1/m) and the normalized power $P_N=\sum_{j} n_j/N=1$ are constant. Extremisation of the optical entropy is solved by a weighted Bose-Einstein law (wBE, see the Section Methods) \cite{zitelli2023statistics}

\begin{equation}
\lvert f_j \rvert ^2=\frac{2(g_j-1)}{g_j\gamma}\frac{1}{\exp\big(-\frac{\mu'+\epsilon_j}{T}\big)-1}    .
\label{eq:BE0}
\end{equation}

In Eq.~\ref{eq:BE0}, $\lvert f_j \rvert^2=2 n_j/(\gamma n_0 g_j)$ is the mean modal power fraction, over two polarizations, in modal group $j$. $\mu'=\mu+\beta_{j=Q}$, with $\mu$ (1/m) the chemical potential and $T$ (1/m) an optical temperature. $\mu'$ and $T$ are two degrees of freedom for fitting Eq. \ref{eq:BE0} to the experimental data, with the only constraint for the normalized power $\gamma$ to scale with the input power, and to respect the conservation law $\sum_{j=1}^{Q}j\lvert f_j \rvert ^2=1$.

Eq. \ref{eq:BE0} has proven capable of reproducing the experimental modal power distributions in nonlinear thermalisation regime, where IM-FWM dominates, and also at lower power, where RMC is mostly responsible for power exchange among modes, at least for long fibre spans. 

The solution Eq. \ref{eq:BE0} eventually applies to all experiments sharing the same normalized values $U_N$ and $P_N$, for example those performed by increasing the input peak power while maintaining a same input distribution. Hence, it is meaningful to compare those experiments by fitting the modal distributions using Eq. \ref{eq:BE0}. In doing so, the validity of the thermodynamic approach is stressed using the error on the state equation (see Section Methods)

\begin{equation}
\epsilon_{SE}=\frac{SE}{\sum_{j=1}^Q \beta_j \frac{g_j}{2} \lvert f_j \rvert^2 -\Big(\mu'-\beta_{j=Q} +(2M-Q) \frac{T}{\gamma}\Big)} .
\label{eq:StateEquationError0}
\end{equation}

In Eq. \ref{eq:StateEquationError0}, $\mu'$ and $T/\gamma$ must scale in order to balance the sum in the first part of the equation, which represents the normalized energy.

Equation \ref{eq:BE0} can be reduced to the well-known RJ law under the assumption $\lvert \mu+\epsilon_j \rvert << \lvert T n_0 \rvert$ (see Section Methods).
The RJ distribution has been widely used in the literature to analyze the modal power distribution in the thermalisation regime \cite{wu2019thermodynamic,Fusaro_PhysRevLett.122.123902,pourbeyram2022direct,Wright:2022NP}. In fact, it is an excellent criterion to recognize thermalised states. However, the assumptions made for its formulation prevent its use at lower power. The wBE is more effective to study modal distributions for variable input power, provided the error on the state equation (Eq. \ref{eq:StateEquationError0}) remains limited to below 5 \%, which is a fair condition to certify the validity of the thermodynamic approach at a given power.


It has been previously reported \cite{baudin2020classical} that, when approaching to the thermalisation regime, the modal distribution approaches to the condensation into the fundamental mode $\lvert f_1 \rvert ^2$; the temperature per unit power $T/\gamma$ ($T'$ in the RJ, see Section Methods) reaches for a minimum value, the chemical potential $-\mu'$ approaches to the eigenvalue of the fundamental mode $\epsilon_1$. The information or Shannon entropy $S_S$ must decrease to a minimum as we approach to the condensation (see Section Methods).
These are typical signs of an achieved thermalisation. In the following sections, it will be provided evidence of thermalisation in self-cleaning experiments in the normal chromatic dispersion region, as well as in multimode soliton condensation experiments in the anomalous dispersion region. Results will be analyzed in terms of Boltzmann and Shannon entropy, and also in terms of temperature and chemical potential, after applying Eq. \ref{eq:BE0} for fitting the experimental data.


\section{Self-cleaning experiment} \label{sec:CleaningExp}

The first set of results come from the re-processing of previously reported experiments in the normal dispersion region \cite{pourbeyram2022direct}. 
Figure 2 of \cite{pourbeyram2022direct} refers to a 0.5 m OM4 GRIN fibre, propagating 200-fs pulses at $\lambda=1040$ nm wavelength. Figure 4 of the same paper refers to a 1.5 m OM4 GRIN fibre, with same input pulses; input peak power was varied from 2.5 to 86.7 kW. The output modal power fraction of $Q=14$ modal groups was extracted by a combined method using both the output near-field (NF) and far-field (FF). 

Fig. \ref{fig:FigSupp_2} of the supplementary material reports  the power fraction of the modes vs. $\epsilon_j$ (red dots), in the experiment of Fig. 2b in \cite{pourbeyram2022direct} at 52 kW input peak power. Statistical power fluctuations are observed for the modal power into each group, which should not be considered as experimental errors; by averaging the modal power fractions into the groups we obtain $\lvert f_j \rvert ^2$ (black circles) which are not affected by statistical fluctuations and can be fitted using the RJ (Eq. \ref{eq:RJ}) or the wBE (Eq. \ref{eq:BE0}); fits are calculated using a standard nonlinear least squares method. In the thermalisation regime, the former law is capable of reproducing the experimental data up to the 8-th modal group (36 modes per polarization), while the latter is accurate up to the 14-th group (105 modes per polarization). From the RJ, we obtain $T'=400$ (1/m), $\mu'=-72626$ (1/m) and the fit R-squared=0.958; from the wBE it is $T=25283$ (1/m), $\mu'=-72388$ (1/m), $T/\gamma=361$ (1/m) and the R-squared=0.996. For both laws, the error on the state equation $\epsilon_{SE}=0.0028$.

\begin{figure}[h]
\includegraphics[width=0.95\textwidth]{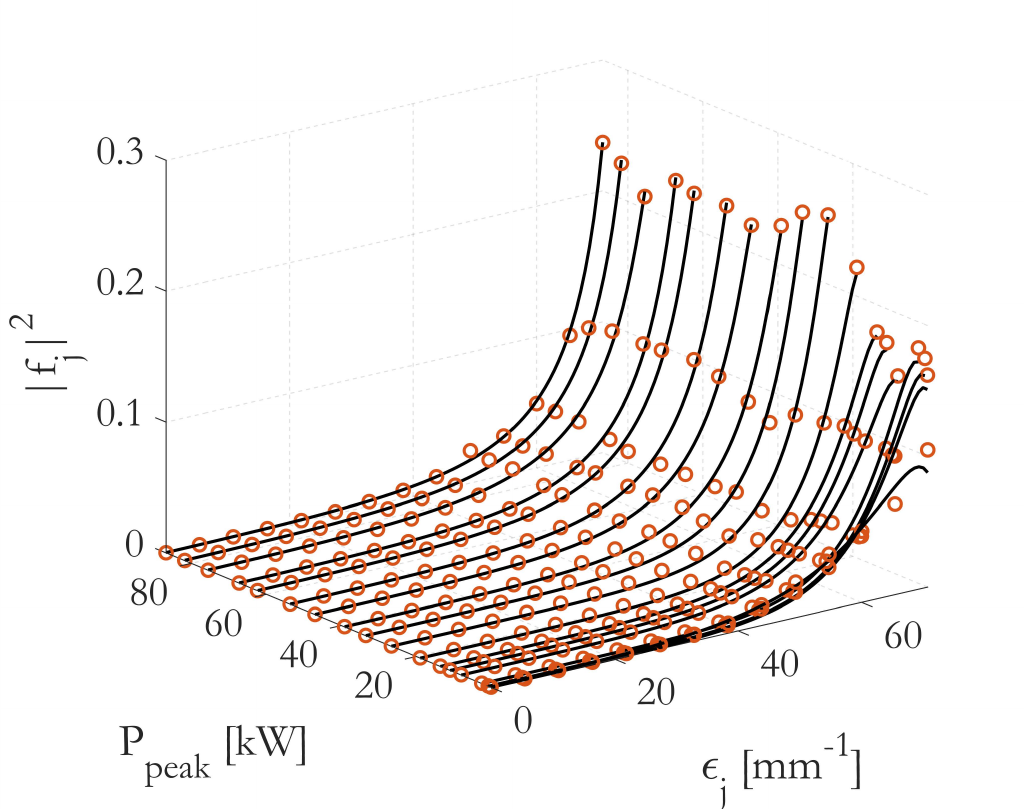}	
\caption{Output mean modal power fraction $\lvert f_j \rvert ^2$ vs. modal group eigenvalues and input peak power. 200-fs pulses were propagated over 1.5 m of GRIN at $\lambda=1040$ nm.}
\label{fig:Fig2_1}
\end{figure}

By proceeding in a similar way with the experiment of Figure 4 in \cite{pourbeyram2022direct}, we obtain the wBE and RJ fits illustrated in Fig. \ref{fig:FigSupp_1} of the supplementary material. Here, it is more evident the larger accuracy obtained using the wBE when reproducing the mean modal power fraction of the HOMs. In Fig. \ref{fig:Fig2_1}, they are resumed the mean modal power fractions $\lvert f_j \rvert ^2$ vs. $\epsilon_j$ and the input peak power $P$. Black curves are the fits performed using Eq. \ref{eq:BE0}. 
Table \ref{tab1} provides the thermodynamic parameters obtained from the RJ and wBE fits. The R-squared values of the wBE fit are always above 0.98, with a maximum of 0.996 at thermalisation power; the R-squared from the RJ range between 0.81 and 0.91 and is greater than 0.90 at 26.6 kW, which indicates the RJ as being accurate only in the thermalisation regime. 

The error on state equation $\epsilon_{SE}$ ranges between 0.210 and 0.008 from the lower to higher power, and is less than 0.05 at 10.9 kW, denoting the validity of the thermodynamic approach from this power level. 

\begin{tiny}
\begin{table}[h]
\caption{Thermodynamic parameters for the experiment of Fig. \ref{fig:Fig2_1}.}
\label{tab1}  
\centering	
\tiny
\begin{tabular*}{1.0\textwidth}{@{}p{0.5cm} p{0.5cm} p{0.7cm} p{0.5cm} p{0.5cm} p{0.7cm} p{0.5cm} p{0.5cm} p{0.5cm} p{0.5cm} p{0.5cm} p{0.5cm} p{0.5cm}@{}}
\toprule
  & RJ & & & wBE & & & & & SE & Entr. & & \\
\midrule
$P$ & $T'$ & $\mu'$ & $R^2$ fit & $T$ & $\mu'$ & $\gamma$ & $T/\gamma$ & $R^2$ fit & $\epsilon_{SE}$ & $S$ & $S_N$ & $S_S$\\
\midrule
 $[$kW] & [m$^{-1}$] & [m$^{-1}$] & & [m$^{-1}$] & [m$^{-1}$] & & [m$^{-1}$] & & & & & \\
\midrule
86.7 & 444 & -72509 & 0.909 & 11904 & -73597 & 15.22 & 782.0 & 0.996 & 0.008 & 4197 & 299 & 3.10\\
81.7 & 462 & -72649 & 0.887 & 11952 & -73856 & 14.35 & 833.2 & 0.996 & 0.008 & 4194 & 317 & 3.11\\
75.5 & 477 & -72866 & 0.862 & 12016 & -74330 & 13.26 & 906.4 & 0.993 & 0.009 & 4177 & 342 & 3.13\\
67.2 & 462 & -72616 & 0.885 & 10875 & -74053 & 11.80 & 921.6 & 0.995 & 0.009 & 4160 & 383 & 3.11\\
62.3 & 463 & -72651 & 0.881 & 10574 & -74234 & 10.94 & 966.7 & 0.992 & 0.010 & 4144 & 412 & 3.12\\
53.7 & 466 & -72654 & 0.876 & 9981 & -74469 & 9.41 & 1060.5 & 0.992 & 0.011 & 4118 & 475 & 3.11\\
47.0 & 461 & -72693 & 0.882 & 9521 & -74836 & 8.25 & 1153.5 & 0.992 & 0.012 & 4109 & 541 & 3.17\\
39.1 & 445 & -72563 & 0.905 & 8607 & -74955 & 6.87 & 1253.7 & 0.988 & 0.013 & 4079 & 646 & 3.18\\
33.5 & 422 & -72353 & 0.931 & 7632 & -74739 & 5.88 & 1297.5 & 0.983 & 0.014 & 4062 & 752 & 3.20\\
26.6 & 430 & -72338 & 0.916 & 7035 & -75017 & 4.67 & 1506.2 & 0.987 & 0.016 & 4050 & 943 & 3.19\\
18.9 & 452 & -72652 & 0.888 & 7117 & -76862 & 3.32 & 2144.7 & 0.984 & 0.023 & 3992 & 1308 & 3.25\\
13.6 & 483 & -73214 & 0.836 & 7858 & -80232 & 2.39 & 3290.8 & 0.989 & 0.036 & 3932 & 1802 & 3.33\\
11.0 & 483 & -73292 & 0.835 & 7834 & -81686 & 1.92 & 4070.7 & 0.990 & 0.046 & 3895 & 2210 & 3.35\\
8.0 & 493 & -73657 & 0.811 & 8337 & -85366 & 1.40 & 5950.1 & 0.991 & 0.069 & 3839 & 2996 & 3.41\\
2.5	 & 481 & -73199 & 0.842 & 7030 & -89444 & 0.44 & 16014.8 & 0.982 & 0.213 & 3604 & 9027 & 3.41\\
\bottomrule
\end{tabular*}
\end{table}
\end{tiny}

\FloatBarrier

\begin{figure}[h]
\includegraphics[width=0.8\textwidth]{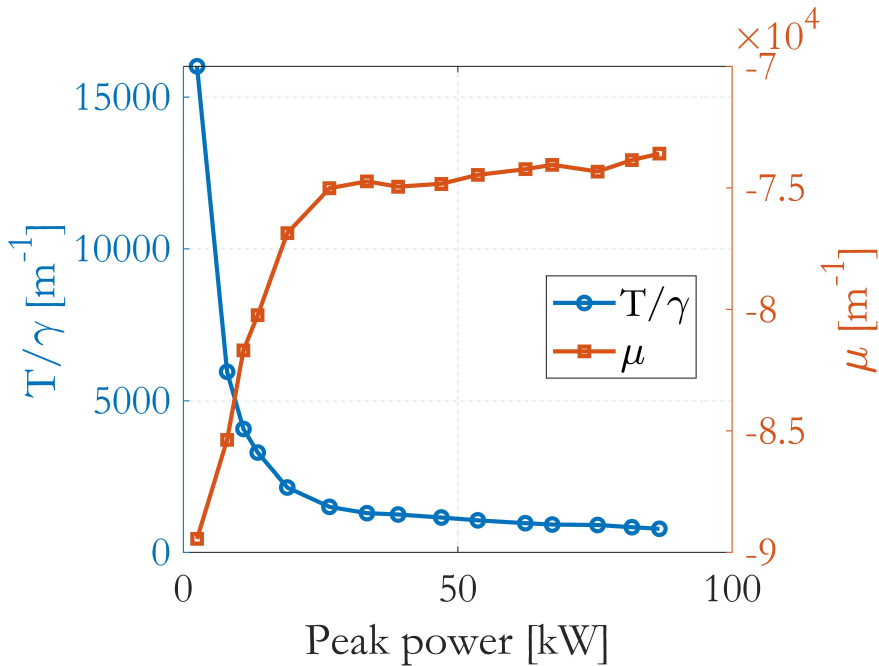}	\centering	
\caption{Temperature $T/\gamma$ and chemical potential $\mu'$ (m$^{-1}$) vs. the input peak power, in the experiment of Fig. \ref{fig:Fig2_1}.}
\label{fig:Fig2_2}
\end{figure}

When comparing the results obtained using the wBE or the RJ, it is clear from Eq. \ref{eq:RJ} that the temperature values at different powers are to be confronted using the temperature per unit power $T/\gamma$ for the wBE or $T'=2T/\gamma$ for the RJ. Hence, starting from the wBE fits, we obtain the curves of Fig. \ref{fig:Fig2_2} for $T/\gamma$ and $\mu'$ (m$^{-1}$). As it was expected, the chemical potential reaches a constant value close to $-\epsilon_1=-70672$ m$^{-1}$ when thermalisation is reached at approximately 27 kW; correspondingly, $T/\gamma$ reduces to a minimum of less than 1500 m$^{-1}$, ten times smaller than at low power.

\FloatBarrier

\begin{figure}[h]
\includegraphics[width=0.8\textwidth]{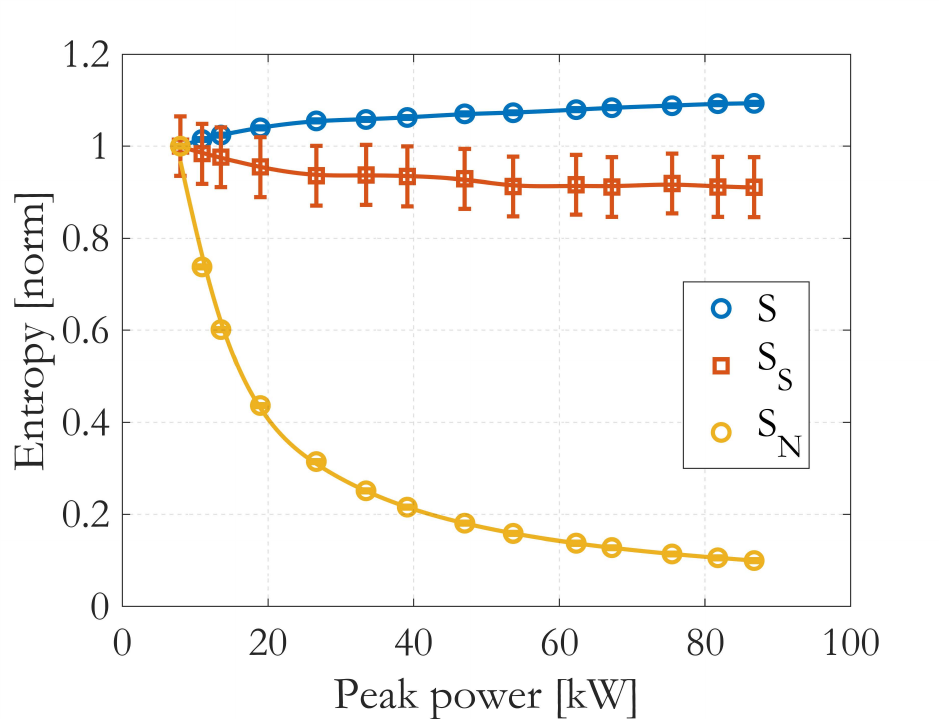}	\centering	
\caption{$S$, $S_N$ and $S_S$, normalized to the values at lowest power, vs. the input peak power in the experiment of Fig. \ref{fig:Fig2_1}.}
\label{fig:Fig2_3}
\end{figure}

The values of the Boltzmann entropy $S$, entropy per unit particle $S_N$ and Shannon entropy $S_S$, as defined in Section Methods, are reported in Fig. \ref{fig:Fig2_3} vs. $P$; entropies are normalized to the values at the minimum power, and errors are reported to demonstrate the effective trends. 

The error on $S$ is calculated as $\Delta S=2\sigma_B$, with $\sigma_B^2=\sum_j g_j^2 \sigma_{nj}^2/n_j^2$; $\sigma_{nj}=\Delta n_j/2$ is the standard deviation for the measurement of $n_j$ or, equivalently, of $\lvert f_j \rvert^2$; in Fig. \ref{fig:Fig2_3}, it was assumed $\Delta n_j/n_j=0.1$, and statistical modal power equipartition was assumed into each group. 
In a similar fashion, the error on $S_N$ is $\Delta S_N=2\sigma_N$, with $\sigma_N^2=\sum_j (g_j-1)^2 \sigma_{nj}^2/n_j^2$.

The error on $S_S$ is calculated as $\Delta S_S=(S_S-1) \Delta (\lvert f_j \rvert^2)/ \lvert f_j \rvert^2$, coming directly from Eq. \ref{eq:ShannonEntropy}; it was assumed $\Delta (\lvert f_j \rvert^2)/ \lvert f_j \rvert^2=0.1$. 

In Fig. \ref{fig:Fig2_3}, the error on $S$ and $S_N$ are much smaller than for $S_S$, because of the $n_j$ factor at the denominator. Despite the large error on $S_S$, it is confirmed its reduction as the thermalisation regime is achieved. $S$, on the contrary, increases with power and reaches a maximum at thermalisation. As expected, $S_N$ tends to zero in the thermalisation regime, despite the increase of the entropy $S$ (see Section Methods).

\section{Multimode solitons experiment} \label{sec:SolitonExp}

The second set of results comes from the re-processing of previously reported experiments in the anomalous dispersion region \cite{zitelli2023statistics}. Figure 2 of \cite{zitelli2023statistics} shows the propagation of 250-fs pulses into 830 m of GRIN OM4 fibre, at $\lambda=1400$ nm, in the anomalous dispersion region of the fibre. Pulses at relatively low power are separated by the modal dispersion, after having interacted by RMC and IM-FWM; at high power, condensation of the pulse train into solitons is achieved.

The formation of trains of quasi-solitons and solitons prevents the pulses from broadening with distance, conserving the validity of the thermodynamic approach, as it is certified by the low values of $\epsilon_{SE}$. The mean modal power fraction was extracted using a 3D modal decomposition method, consisting of sampling the peak powers of the train of pulses, after measurement with a fast photodiode. The near-field is then reconstructed and compared with the one measured by an infrared camera \cite{Zitelli_23}.

\begin{figure}[h]
\includegraphics[width=0.95\textwidth]{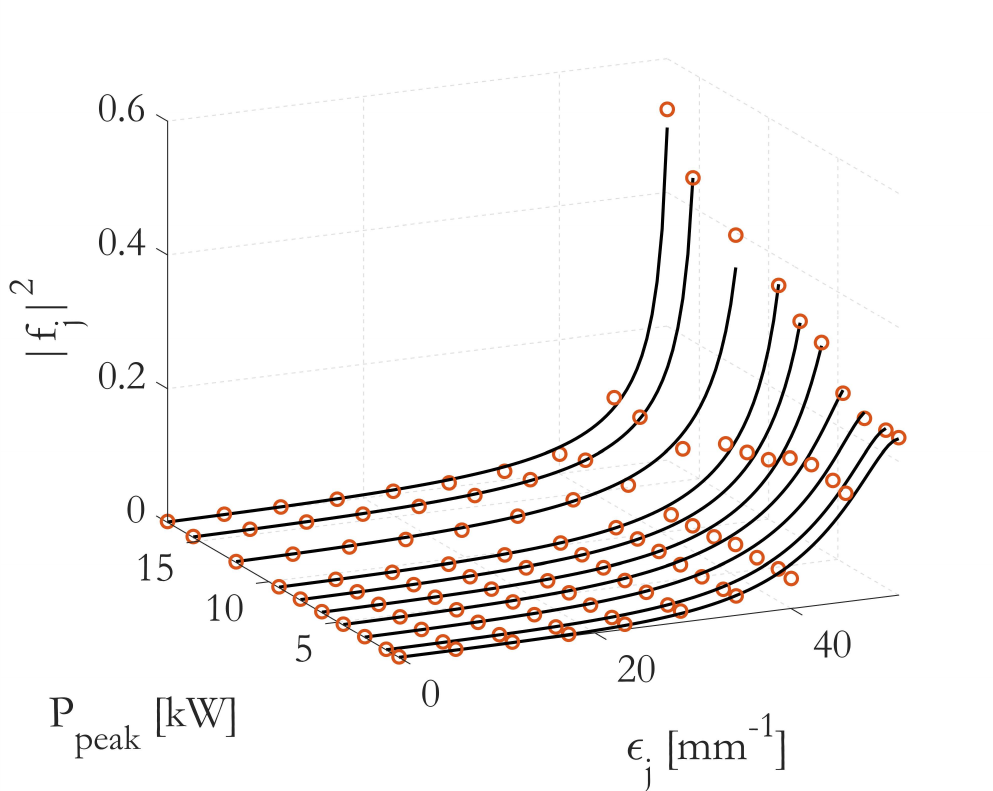}	\centering	
\caption{Output mean modal power fraction $\lvert f_j \rvert ^2$ vs. modal group eigenvalues and input peak power. 250-fs pulses were propagated over 830 m of GRIN at $\lambda=1400$ nm.}
\label{fig:Fig3_1}
\end{figure}

The measured mean modal power fraction, for input peak power ranging between 0.62 kW and 17.3 kW and for $Q=10$ modal groups, is reported in Fig. \ref{fig:Fig3_1}. Also in this case, black curves are the wBE fits, showing good accuracy at all powers; for comparison, the RJ fits are reported in Fig. \ref{fig:FigSupp_1} of the supplementary material. 

Table \ref{tab2} reports the corresponding thermodynamic parameters. The RJ fails to follow the experimental data below 7.7 kW; with regard to the wBE, the R-squared ranges between 0.957 to 0.999 from lower to higher power. 

The error on the state equation $\epsilon_{SE}$ ranges between 0.075 and 0.003, and is lower than 0.05 already at $P=1.55$ kW, denoting the validity of the thermodynamic approach from this power level.

\begin{table}[h]
\caption{Thermodynamic parameters for the experiment of Fig. \ref{fig:Fig3_1}.}
\label{tab2}
\centering
\tiny
\begin{tabular*}{1.0\textwidth}{@{}p{0.5cm} p{0.5cm} p{0.7cm} p{0.5cm} p{0.5cm} p{0.7cm} p{0.5cm} p{0.5cm} p{0.5cm} p{0.5cm} p{0.5cm} p{0.5cm} p{0.5cm}@{}}
\toprule
  & RJ & & & wBE & & & & & SE & Entr. & & \\
\midrule
$P$ & $T'$ & $\mu'$ & $R^2$ fit & $T$ & $\mu'$ & $\gamma$ & $T/\gamma$ & $R^2$ fit & $\epsilon_{SE}$ & $S$ & $S_N$ & $S_S$\\
\midrule
 $[$kW] & [m$^{-1}$] & [m$^{-1}$] & & [m$^{-1}$] & [m$^{-1}$] & & [m$^{-1}$] & & & & & \\
\midrule
17.31 & 388.8 & -51775 & 0.979 & 10593 & -52004 & 20.10 & 527 & 0.999 & 0.003 & 1851 & 92.1 & 2.18\\
15.46 & 418.2 & -51934 & 0.976 & 10571 & -52240 & 18.00 & 587 & 0.999 & 0.004 & 1845 & 102.5 & 2.35\\
12.36 & 525.9 & -52501 & 0.918 & 11240 & -53041 & 14.40 & 781 & 0.983 & 0.005 & 1823 & 126.6 & 2.53\\
9.27 & 511.2 & -52408 & 0.929 & 9530 & -53174 & 10.81 & 882 & 0.990 & 0.006 & 1798 & 166.3 & 2.51\\
7.73 & 538.2 & -52660 & 0.910 & 9274 & -53749 & 9.02 & 1029 & 0.988 & 0.007 & 1790 & 198.5 & 2.63\\
6.18 & 559.1 & -52808 & 0.891 & 8790 & -54281 & 7.22 & 1217 & 0.985 & 0.008 & 1762 & 244.0 & 2.64\\
4.64 & 626.3 & -53414 & 0.809 & 8980 & -55873 & 5.38 & 1668 & 0.968 & 0.012 & 1732 & 321.7 & 2.68\\
3.09 & 647.5 & -53652 & 0.770 & 8337 & -57361 & 3.59 & 2323 & 0.957 & 0.017 & 1690 & 470.8 & 2.69\\
1.55 & 655.3 & -53736 & 0.756 & 7381 & -59891 & 1.79 & 4113 & 0.959 & 0.031 & 1620 & 902.6 & 2.69\\
0.62 & 651.0 & -53756 & 0.758 & 6701 & -64003 & 0.72 & 9335 & 0.961 & 0.075 & 1533 & 2135.9 & 2.72\\
\bottomrule
\end{tabular*}
\end{table}

\FloatBarrier

\begin{figure}[h]
\includegraphics[width=0.8\textwidth]{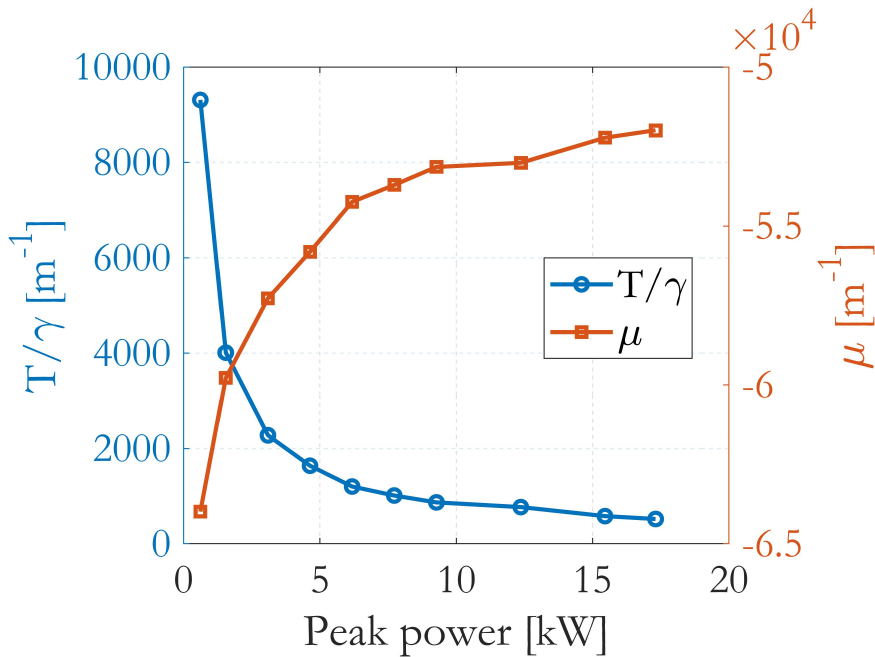}	\centering	
\caption{Temperature $T/\gamma$ and chemical potential $\mu$ (m$^{-1}$) vs. the input peak power, in the experiment of Fig. \ref{fig:Fig3_1}.}
\label{fig:Fig3_2}
\end{figure}

Fig. \ref{fig:Fig3_1} shows the approaching to a condensed state, characterized by the content of the fundamental mode $\lvert f_1 \rvert^2$ which increases from 0.23 to 0.53. Correspondingly, in Fig. \ref{fig:Fig3_2} the chemical potential $\mu'$ obtained from the wBE approaches to $-\epsilon_1=-50994$ m$^{-1}$, and $T/\gamma$ to less than 890 m$^{-1}$ for power larger than 10 kW, ten times smaller than at low power.

\begin{figure}[h]
\includegraphics[width=0.8\textwidth]{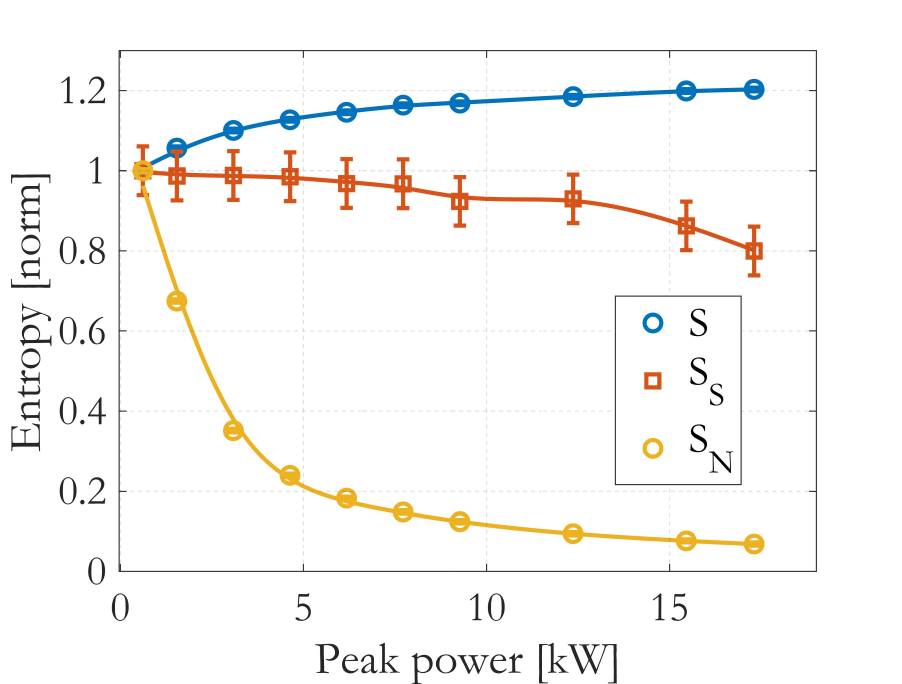}	\centering	
\caption{$S$, $S_N$ and $S_S$, normalized to the values at lowest power, vs. the input peak power in the experiment of Fig. \ref{fig:Fig3_1}.}
\label{fig:Fig3_3}
\end{figure}

The parameter $S_N$ and the entropy $S_S$ in Fig. \ref{fig:Fig3_3} experience a drop at high power, when approaching to soliton condensation. The entropy $S$, on the contrary, increases with power and converges to a maximum. 
Also in this case, like in the self-cleaning experiment, $S_N$ and $S_S$ reach for a minimum of information when condensation takes place.

\FloatBarrier

\section{Discussion and conclusions}

A new thermodynamic theory of multimode systems has been expanded respect to \cite{zitelli2023statistics}, introducing new formulae for the optical entropy and the fundamental thermodynamic equation. 
The wBE law, Eq. \ref{eq:BE0}, has demonstrated good accuracy in reproducing the mean modal power distribution not only in the nonlinear thermalisation regime, where IM-FWM produces inter-modal mixing, but also at lower power when RMC dominates. 

Experimental data from \cite{pourbeyram2022direct} and \cite{zitelli2023statistics} have been post-processed in terms of the new theory, finding interesting analogies between the self-cleaning process in the normal chromatic dispersion region of a GRIN fibre and the soliton condensation regime in the anomalous dispersion region.

In the self cleaning experiment, IM-FWM produced a power transfer to the lower-order groups for increasing pulse peak power; population of the fundamental mode converged to more than 22\% for $P > 19$ kW, starting from 10\% at low power. The temperature per unit power $T/\gamma$, calculated using the wBE, dropped by a factor 520\% for power values increasing from 11 to 86.7 kW (values where the error $\epsilon_{SE} < 0.05$); correspondingly, the chemical potential $\mu'$ approached to the eigenvalue of the fundamental mode $-\epsilon_1=-70670$ (1/m). These are clear indications of a reached thermalisation state, confirmed by an increase of the Boltzmann entropy $S$ by a factor 16\%. The information entropy $S_S$  and the entropy per unit particle $S_N$, on the contrary, decreased by a factor 10\% and 90\%, respectively. 

It was demonstrated in the Section Methods that a decrease of $S_N$ following an increase of $T$, $\gamma$ and $S$ is possible under the conditions imposed by Eqs. \ref{eq:EntropyDiff2} and \ref{eq:EntropyDiff3}; in the experiment, those conditions were satisfied by modes belonging to groups with $j \geq 4$, which have larger weight to the entropy changes. The Shannon entropy $S_S$, on the other hand, is related to the modal occupation instead of the number of microstates; as the particles condensate to the fundamental mode, we expect to find a minimum of information in terms of modal distribution.

In the soliton condensation experiment, a complex interplay of modal dispersion, RMC and IM-FWM results into the generation of a single pulse or a train of pulses propagating as quasi-solitons or solitons, depending on the input power and coupling conditions \cite{zitelli2023statistics}. In the absence of RMC, pulses carrying the individual modal groups would be lacking of the modes from other groups or the fundamental mode. RMC generates a background noise composed by all modes which contribute to populate the propagating pulses. IM-FWM therefore transfers power to the lower order modes present in all pulses, generating a single soliton or a train of condensed solitons. In this context, it was observed a modal power re-distribution for increasing $P$, with the power fraction of the fundamental mode passing from 23\% to 52\% from the lower to the higher power. $T/\gamma$ decreased by a factor 780\% when passing from 1.55 to 17.3 kW (corresponding to $\epsilon_{SE} < 0.05$). In the same power range, the $\mu'$ increased approaching to $-\epsilon_1=-51000$ (1/m), the entropy S increased by a factor 14\%, $S_S$ and $S_N$ decreased of 19\% and 90\%, respectively.

The operative difference between the two regimes is principally the fibre length. In self cleaning experiments, the most detrimental effect for the validity of the thermodynamic approach is chromatic dispersion, which can be considered as a dissipative effect itself; pulse broadens by a factor $\sqrt{2}$ after a dispersion length $L_D=T_0^2/\beta_2$, being $T_0$ the pulse width and $\beta_2$ (s$^2$/m) the fibre dispersion; the peak power $P$ and peak number of particles $N$ decreases accordingly, producing an error $\epsilon_{SE}$ which may go well above 5\%. fibre length is therefore limited to few meters and, in some cases, less than 1 m.

In soliton condensation experiments, RMC is a necessary step to produce a single condensed pulse or a train of pulses. Pulse broadening is compensated by self-phase modulation (SPM), hence, the chromatic dispersion does not behave as a dissipative effect. Long distances (hundred of meters) are necessary to observe a power transfer to the fundamental mode, as a consequence of the interplay between modal dispersion, RMC and IM-FWM. RMC generates a limited change to the internal energy because of particle transfer to other groups; if, for example, a complete transfer of power is produced by RMC from the modal group $j=10$ to $j=1$ the energy change is $(\beta_{j=10}-\beta_{j=1})/\beta_{j=10} \leq 0.008$. For fully formed solitons, at power higher that those considered here, inter-modal stimulated Raman scattering (IM-SRS) \cite{Gordon1986}\cite{Zitelli_9887813} is responsible for wavelength red-shift of the formed solitons; in this case, a wavelength shift $\lambda_2-\lambda_1=80$ nm produces a change to the internal energy of $[\beta(\lambda_2)-\beta(\lambda_1)]/\beta(\lambda_1) \leq 0.05$. Hence, even Raman nonlinearity and RMC can be compatible with the thermodynamic approach, and the error $\epsilon_{SE}$ must be used as an overall check for the thermodynamic validity of the experiment.

The surprising similarities between the results of the two experiments, which involve all the thermodynamic parameters, indicate a common basis for condensation processes in the normal and anomalous chromatic dispersion regimes. Beam self-cleaning and soliton condensation are both caused by IM-FWM as the primary process, and both are observed at $P$ ranging between 10 to 30 kW. Thermodynamics based on wBE has proven to be a powerful tool for the analysis of processes apparently different from an operational point of view, but similar in thermodynamic terms.

\section{Methods} \label{sec:Methods}

\subsection{Weighted Bose-Einstein and Raileigh-Jeans laws} \label{subsec:wBE_RJ}

In the main text, it was considered an optical multimode system including $Q$ groups of degenerate modes, distributed over $g_j/2$ modes and 2 polarizations, with $j=1, 2, .., Q$; hence, $g_j$ is the degeneracy over two polarizations; in the special case of a GRIN fibre, it is $g_j=2, 4, 6,.., 2Q$; the number of modes and polarizations is $2M=Q(Q+1)$. 

 The nearly degenerate modes of a group have substantially equal propagation constant \cite{Olshansky:75} $\beta_j =n_{co} k_0 \sqrt{1-2\Delta (j/Q)}$ (m$^{-1}$), with $n_{co}$ the core index and $\Delta=(n_{co}^2-n_{cl}^2)/2n_{co}^2$ the relative core-cladding index difference.

Suppose that $n_j$ is the bosonic population into modal group $j$, distributed over $g_j$ nearly-degenerate modes and polarizations. $\epsilon_j=\beta_j-\beta_{j=Q}$ are the differential modal eigenvalues. The total number of particles in the system $N=\sum_{j=1}^Q n_j$. Power can be written as $P=\sum_{j=1}^Q n_j P_0/n_0$ (in units of W); the system's internal energy is $U=-\sum_{j=1}^Q \beta_j n_j P_0/n_0$ (in units of W/m). 
$\gamma=N/n_0=P/P_0$ is the total fractional power, with $n_0$ a reference number of particles and $P_0$ the corresponding input peak power. The multiplicity of the system, as the number of possible microstates populated by the particles, is given by

\begin{equation}
W=\prod_{j=1}^Q \frac{(n_j+g_j-1)!}{n_j!(g_j-1)!}   .
\label{eq:Multiplicity}
\end{equation}

The Boltzmann entropy of the system $S=\ln(W)$ is related to the number of microstates; it reads \cite{wu2019thermodynamic}

\begin{equation}
S=\sum_{j=1}^{Q}(n_j+g_j-1)\big[\ln(n_j+g_j-1)-1\big]-(g_j-1)\big[\ln(g_j-1)-1\big]-n_j\big[\ln(n_j)-1\big]    .
\label{eq:Entropy}
\end{equation}

The quantity $S_N=S/\gamma=n_0\ln{(W)}/N$ has the meaning of an entropy per unit particle; although it cannot be considered as an entropy itself (for example, it is not additive), we may try to find an extremum of $S_N$ while assuming constant the system's normalized internal energy $U_N=U/P=-\sum_{j}\beta_j n_j/N$ (1/m) and the power $P_N=P/P=\sum_{j} n_j/N=1$ (both are normalized to the power $P=\gamma P_0=NP_0/n_0$)
 
 \begin{equation}
\frac{\partial}{\partial n_l}\Big[\ln{(W)}/N+ \sum_{j=1}^Q \Big(a n_j/N +b \beta_j n_j/N \Big)\Big]=0      .
\label{eq:EntropyDerivative1}
\end{equation}

 Eq. \ref{eq:EntropyDerivative1} is multiplied by $N$ and solved using two possible sets of Lagrange multipliers $(a, b)$ or $(a', b')$ \cite{zitelli2023statistics}. A first solution, related to the choice $(a, b)$, is the weighted Bose-Einstein law (wBE) 
 
\begin{equation}
\lvert f_j \rvert ^2=\frac{2(g_j-1)}{g_j\gamma}\frac{1}{\exp\big(-\frac{\mu'+\epsilon_j}{T}\big)-1}    ;
\label{eq:BE}
\end{equation}

in Eq. \ref{eq:BE}, $\lvert f_j \rvert^2=2 n_j/(\gamma n_0 g_j)$ is the mean modal power fraction, over two polarizations, in modal group $j$. $\mu'=\mu+\beta_{j=Q}$, with $\mu$ (1/m) a chemical potential and $T$ (1/m) an optical temperature. 
$\mu'$ and $T$ are two degrees of freedom for fitting Eq. \ref{eq:BE} to the experimental data. The $\gamma$ parameter is free at only one intermediate power; for other powers, it must scale proportionally to $P$ or $N$; the second constraint is the respect of the conservation law $\sum_{j=1}^{Q}(g_j/2)\lvert f_j \rvert ^2=1$.

Hence, the wBE distribution is valid in experiments where $U_N=const$, or equivalently, where $\gamma$ and $U$ scale proportionally or they are constant.

A second solution, related to the choice $(a', b')$, is obtained under the assumption $\lvert \mu'+\epsilon_i \rvert << \lvert T n_0 \rvert$, which leads to the Rayleigh-Jeans (RJ) distribution:

\begin{equation}
\lvert f_j \rvert ^2=-\frac{2(g_j-1)}{g_j \gamma}\frac{T}{\mu'+\epsilon_j} \simeq -\frac{T'}{\mu'+\epsilon_j}    ,
\label{eq:RJ}
\end{equation}

with $T'=2T/\gamma$ (1/m). In Eq. \ref{eq:RJ}, the power factor $\gamma$ was included into the temperature $T'$; hence, $T'$ has meaning similar to the parameter $T/\gamma$ of the wBE, when comparing the results from Eq. \ref{eq:BE} and Eq. \ref{eq:RJ}.
The two sets of Lagrange multipliers used to obtain Eqs. \ref{eq:BE} and \ref{eq:RJ} bring to the same values of internal energy and power, namely \cite{zitelli2023statistics}

\begin{equation}
\sum_{j=1}^Q (a' + b' \beta_j) n_j=\sum_{j=1}^Q a n_j + b \beta_j n_j=Q-2M=\frac{\mu N}{T n_0}+\frac{1}{T n_0}\Big(-\frac{UN}{P}\Big)  ,
\label{eq:StateEquation1}
\end{equation}

\noindent which provides a common state equation (SE)

\begin{equation}
U-\mu P=(2M-Q)P_0T  .
\label{eq:StateEquation2}
\end{equation}

The SE can be rewritten in terms of $U_N=U/P$, $\gamma$ and fitting parameters $\mu'$ and $T$

\begin{equation}
SE=-U_N+\mu+V\frac{T}{\gamma}=\sum_{j=1}^Q \beta_j \frac{g_j}{2}\lvert f_j \rvert^2 +\mu'-\beta_{j=Q} +V \frac{T}{\gamma}=0 ,
\label{eq:StateEquation3}
\end{equation}

where we introduced $V=(2M-Q)$ as the system volume. The experimental error on the SE can be calculated as

\begin{equation}
\epsilon_{SE}=\frac{SE}{\sum_{j=1}^Q \beta_j \frac{g_j}{2} \lvert f_j \rvert^2 -\Big(\mu'-\beta_{j=Q} +V \frac{T}{\gamma}\Big)} .
\label{eq:StateEquationError}
\end{equation}

The error $\epsilon_{SE}$ can be used to certify the validity of the thermodynamic approach. Experimental data show that an error of less than 1\% is obtained when approaching to the system thermalisation or condensation.

\subsection{Optical entropy} \label{subsec:OpticalEntropy}

The Boltzmann entropy (Eq. \ref{eq:Entropy}) can be approximated for $n_j >> g_j$ to

\begin{equation}
S=\sum_{j=1}^{Q}(g_j-1)\ln(n_j) .
\label{eq:EntropyB}
\end{equation}

Using Eqs. \ref{eq:BE} and \ref{eq:StateEquation2} for the wBE and the SE, we get

\begin{equation}
S=-\sum_{j=1}^{Q}(g_j-1) \ln \Big[\frac{\exp \big(-\frac{\beta_j}{T}-\frac{U_N}{T}+\frac{V}{\gamma} \big)-1}{n_0 (g_j-1)} \Big] .
\label{eq:EntropyB2}
\end{equation}

By considering the normalized internal energy $U_N$, number of particles $\gamma$ and volume $V$ as dependent on $T$, derivatives are calculated as $(\partial S/\partial X)_Y=\partial S/\partial X+(\partial S/\partial T)(\partial T/\partial X)$, with $X,Y=U_N, \gamma, V$ or $T$, obtaining the  fundamental equation

\begin{equation}
dS=\sum_{j=1}^{Q} q_j \Big[\frac{\gamma}{T}dU_N-\frac{\mu}{T}d\gamma-dV-\gamma\Big(\frac{\beta_j+U_N}{T^2}\Big)\Big(\frac{\partial T}{\partial U_N}dU_N+ \frac{\partial T}{\partial \gamma}d\gamma+ \frac{\partial T}{\partial V}dV+dT \Big)\Big]  ,
\label{eq:EntropyDiff}
\end{equation}

with weight factor

\begin{equation}
q_j=\frac{g_j}{2} \lvert f_j \rvert^2 \exp \Big( -\frac{\mu'+\epsilon_j}{T} \Big)  .
\label{eq:EntropyPar}
\end{equation}

In Eq. \ref{eq:EntropyPar}, the modal group fractions multiply the weight $\exp[-(\mu'+\epsilon_j)/T]$, which in multimode fibres is larger than 500 for modal group $j=14$ and approaches to less than 10 for the fundamental mode; hence, populated HOMs mostly contribute to the entropy changes. It is therefore important to use fitting laws which accurately describe also the HOM content, like the wBE. 

In experiments involving proportional changes of $U$ and $\gamma$, like those considered in this work, the normalized energy $U_N=U/\gamma P_0$ is constant (the normalized number of particles is always unity). In this case, we must refer to the quantity $S_N=S/\gamma$ which, although non representing an entropy, is extremised by Eq. \ref{eq:EntropyDerivative1} under the constraint $U_N=const$ and brings in any case to the wBE and RJ. Therefore, for the experiments considered in this work, it is meaningful to calculate this quantity too. After derivatives with respect to $U_N$, $\gamma$, $T$ and $V=2M-Q$ we obtain

\begin{equation}
dS_N=\sum_{j=1}^{Q} q_j \Big[\frac{dU_N}{T}+\frac{U_N-\mu}{\gamma T}d\gamma -\frac{dV}{\gamma} -\frac{\beta_j+U_N}{T^2} \Big(\frac{\partial T}{\partial U_N}dU_N+ \frac{\partial T}{\partial \gamma}d\gamma+ \frac{\partial T}{\partial V}dV+dT \Big) \Big]  .
\label{eq:EntropyNormDiff}
\end{equation}

In Eqs. \ref{eq:EntropyDiff} and \ref{eq:EntropyNormDiff} we can consider the volume as a constant ($dV=0$); in fact, the number of supported modes depend on the multimode architecture and not on the input power or modal distribution. Alternatively, the number of modes $M$ and groups $Q$ is dependent on the experimental accuracy for the modal fraction of the HOMs; it is reasonable to limit the number of groups $Q$ to those measurable within a given accuracy.

In those experiments where the non-normalized energy and power are constant ($dU=d\gamma=0$) we find from Eq. \ref{eq:EntropyDiff} that $dS > 0$ for $-\gamma(\beta_j+U_N)dT/T^2 > 0$. Hence, by using Eq. \ref{eq:StateEquation2}, an increase of the entropy is obtained following an increase of the temperature ($dT > 0$) for

\begin{equation}
\beta_j < -U_N  .
\label{eq:EntropyDiff2}
\end{equation}

Since populated HOMs have much larger weight in Eq. \ref{eq:EntropyDiff}, it is important that Eq. \ref{eq:EntropyDiff2} is satisfied for the populated HOMs and not necessarily for the lower-order groups. 

In the experiments considered in this work, it is $dU_N=dV=0$. In this case, from Eq. \ref{eq:EntropyDiff}
we have an increase of entropy ($dS > 0$) for 

\begin{equation}
-\frac{\mu}{T}d\gamma > \gamma \frac{\beta_j+U_N}{T^2} \big( \frac{\partial T}{\partial \gamma}d\gamma+dT \big)  ;
\label{eq:EntropyDiff3}
\end{equation}

the first term in Eq. \ref{eq:EntropyDiff3} is always positive for $d\gamma > 0$; if we also assume $dT > 0$, Eq. \ref{eq:EntropyDiff3} is valid whenever Eq. \ref{eq:EntropyDiff2} is satisfied at least for the HOMs with higher weight.
For example, in the self-cleaning experiments analyzed in this work (Fig. \ref{fig:Fig2_1}), Eq. \ref{eq:EntropyDiff2} is satisfied for group orders with $j \geq 4$.

We want to demonstrate that it is possible to obtain a decrease of the entropy per unit particle $dS_N < 0$ after an increase of the entropy ($dS > 0$) and of the temperature with the input power ($dT/d\gamma > 0$). From Eq. \ref{eq:EntropyNormDiff}, assuming $dU_N=dV=0$ 
and using again Eq. \ref{eq:StateEquation2}, a decrease of $S_N$ is possible for

\begin{equation}
\frac{dT}{d \gamma} > \frac{T^2 V}{\gamma^2 (\beta_j+U_N)}  ;
\label{eq:EntropyNormDiff2}
\end{equation}

if Eqs. \ref{eq:EntropyDiff2} is satisfied, it is $dS > 0$ from Eq. \ref{eq:EntropyDiff3};
also, the second term in Eq. \ref{eq:EntropyNormDiff2} is negative, and the equation is always satisfied for $dT/d\gamma > 0$.

We are also interested in considering the information or Shannon entropy $S_S$, related to the classical concept of modal occupation instead of microstates distribution. In this case we must refer to the $M$ modes with population $m_i=N\lvert f_i \rvert^2$, $i=1,2,..,M$; the modes are not distinguished by polarization, and it is still $2M=Q(Q+1)$. The system multiplicity is reformulated in this case as \cite{Jaines2003}

\begin{equation}
W_S= \frac{N!}{\prod_{i=1}^M m_i!} ,
\label{eq:MultiplicityShannon}
\end{equation}

and the Shannon entropy is calculated per unit particle as

\begin{multline}
S_S= \frac{\ln(W_S)}{N}=\frac{1}{N}\big[\ln(N!)-\sum_{i=1}^M \ln(m_i!)\big] \simeq \frac{1}{N} \big[N\ln(N)-\sum_{i=1}^M N\lvert f_i \rvert^2 \ln(N\lvert f_i \rvert^2)\big] = \\
= -\sum_{i=1}^M \lvert f_i \rvert^2 \ln(\lvert f_i \rvert^2)    ;
\label{eq:ShannonEntropy}
\end{multline}

in Eq. \ref{eq:ShannonEntropy}, it was applied the Stirling's approximation, valid for $m_i >> 1$, and also $\sum_{i=1}^M \lvert f_i \rvert^2 =1$. For systems with degeneracy, assuming statistical modal power equipartition into the groups, it is $\lvert f_j \rvert=\lvert f_i \rvert$ for groups with $g_j$ modes and polarizations, and mode index $i$ corresponding to group index $j$. Namely, for a GRIN fibre, it is $g_j=2$ and $i=1$ for $j=1$, $g_j=4$ and $i=\{1, 2\}$ for $j=2$, up to $g_j=2Q$ and $i=\{j(j+1)/2)-j+1,..,j(j+1)/2\}$ for $j=Q$. The Shannon entropy can be rewritten in this case as

\begin{equation}
S_S= -\sum_{j=1}^Q \frac{g_j}{2} \lvert f_j \rvert^2 \ln(\lvert f_j \rvert^2)    .
\label{eq:ShannonEntropy2}
\end{equation}

\newpage

\section{Supplementary Material}

\begin{figure}[h]
\includegraphics[width=0.8\textwidth]{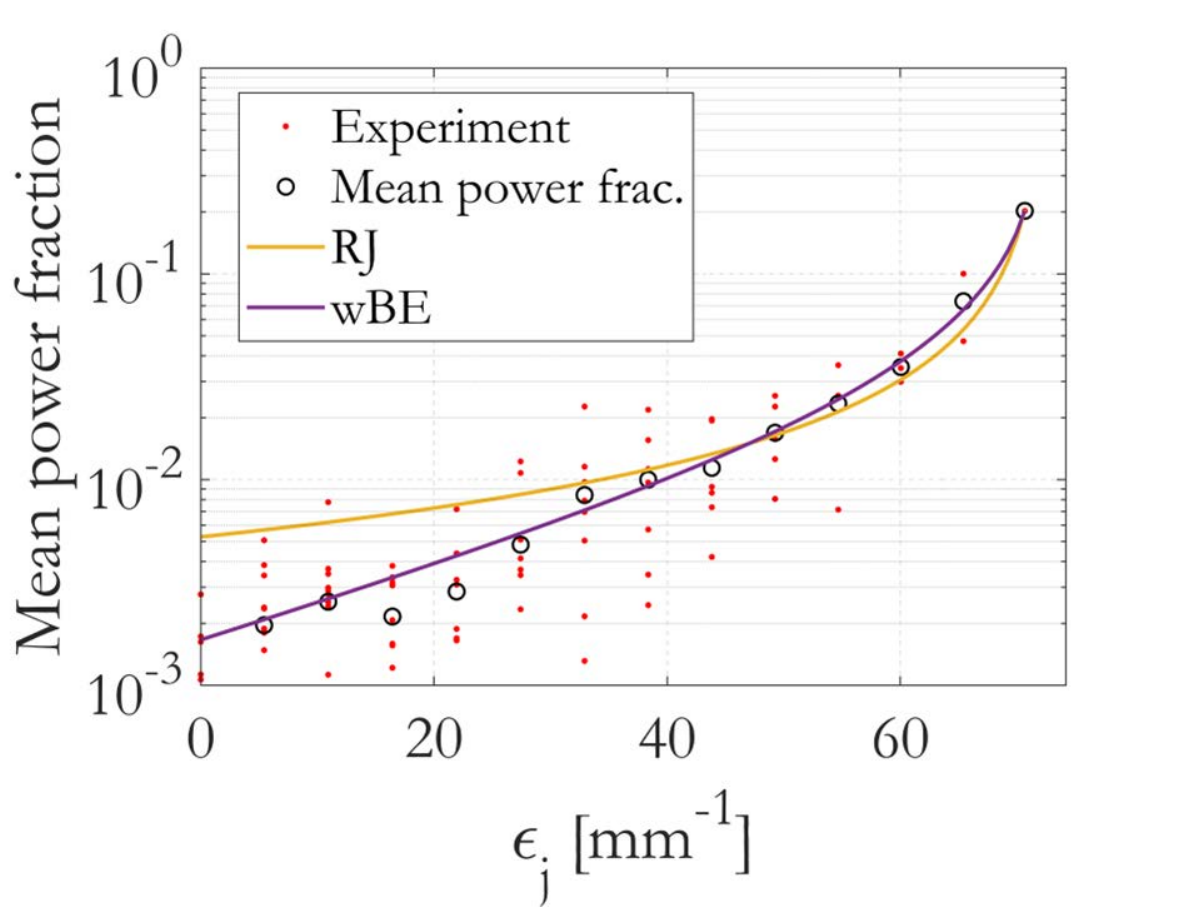}	
\centering	
\caption{Modal power fractions (red dots) and mean modal power fractions (black circles) of the modes vs. eigenvalues $\epsilon_j$, in the experiment of Fig. 2 of \cite{pourbeyram2022direct}, propagating 200 fs input pulses over 50 cm of GRIN OM4 fibre, at 1040 nm and 52 kW input peak power. RJ and wBE best fits of the mean power fractions are calculated.}
\label{fig:FigSupp_2}
\end{figure}
\FloatBarrier

\begin{figure}[h]
\includegraphics[width=0.95\textwidth]{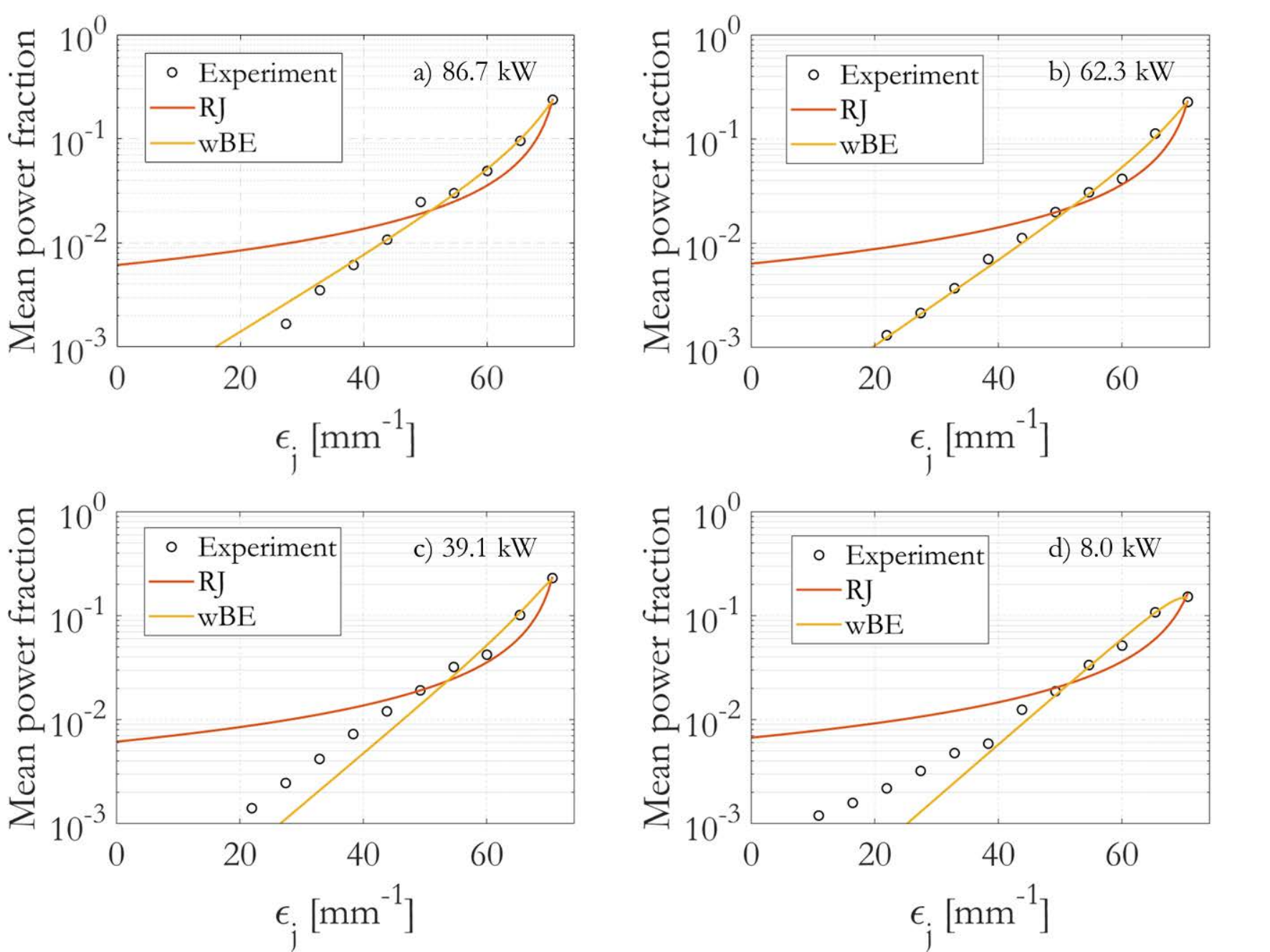}	
\centering	
\caption{Mean modal power fractions (black circles) of the modes vs. eigenvalues $\epsilon_j$, in the experiment of Fig. 4 of \cite{pourbeyram2022direct}, propagating 200 fs input pulses over 150 cm of GRIN OM4 fibre, at 1040 nm and variable power. RJ and wBE best fits of the mean power fractions are calculated.}
\label{fig:FigSupp_1}
\end{figure}
\FloatBarrier

\begin{figure}[h]
\includegraphics[width=0.95\textwidth]{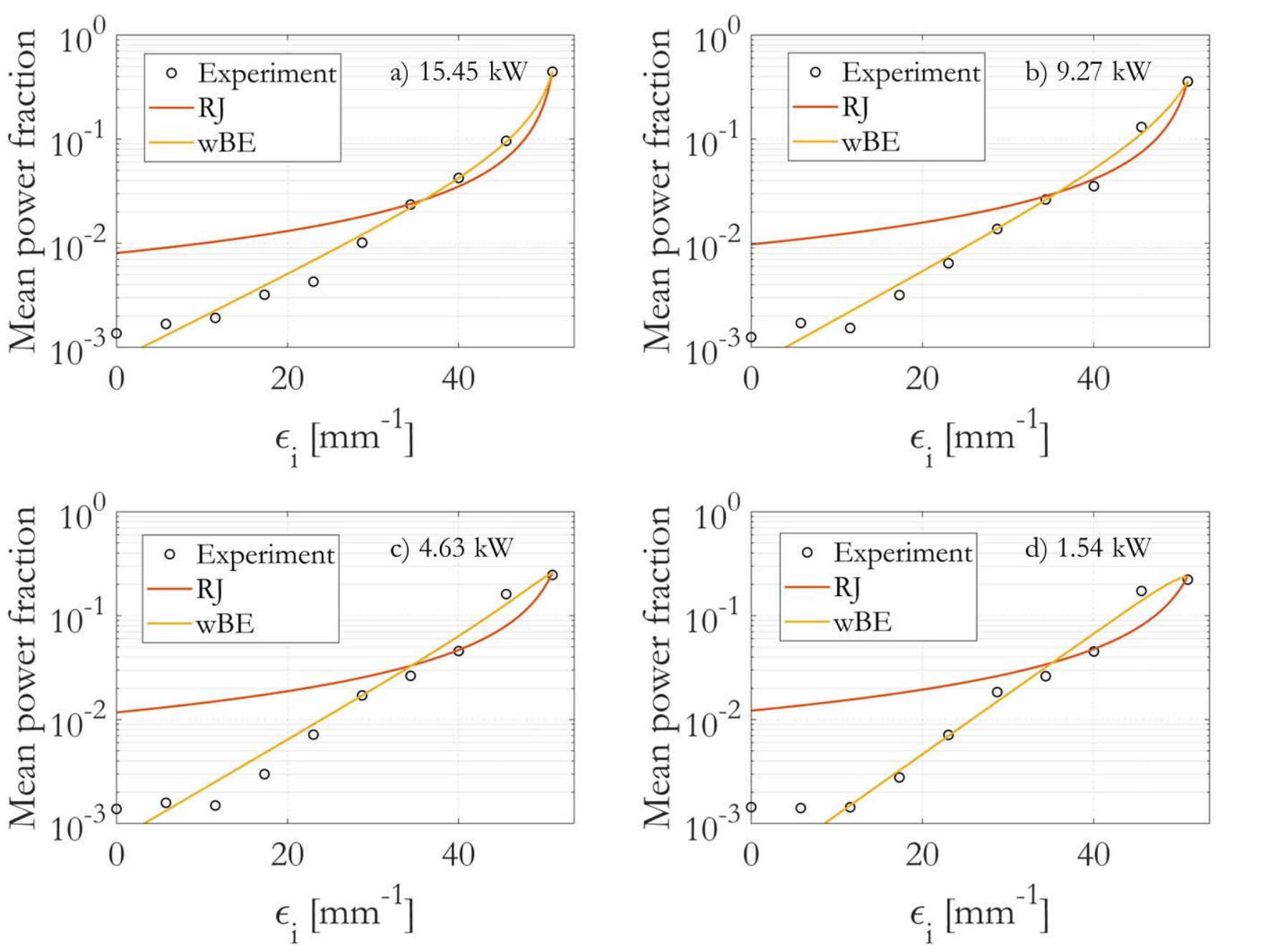}	
\centering	
\caption{Mean modal power fractions vs. eigenvalues $\epsilon_j$ in the experiment of Figure 2 of \cite{zitelli2023statistics}, consisting in soliton propagation over 830 m of GRIN OM4 fibre from 250-fs input pulses at $\lambda=1400$ nm, in the anomalous dispersion region of the fibre. RJ and wBE best fits of the mean power fractions are calculated.}
\label{fig:FigSupp_3}
\end{figure}
\FloatBarrier

\newpage

\backmatter




\section*{Declarations}


\begin{itemize}
\item Acknowledgments

The author wishes to thank F. Wise, H. Pourbeyram and F. Wu for fruitful discussions and for making freely available the self-cleaning data processed in this work.

\item Funding

Project ECS 0000024 Rome Technopole, - CUP B83C22002820006, Funded by the European Union - NextGenerationEU.

\item Conflict of interest/Competing interests 

The author declares no conflict of interest.



\end{itemize}

\noindent











\end{document}